\documentclass[aps,jcp,reprint]{revtex4-1}

\usepackage{amsmath}
\usepackage{graphicx}
\usepackage{color}

\DeclareMathOperator{\Tr}{Tr}

\begin{document}

%\title{Shear and bulk viscosity of a DPD fluid interacting via random and dissipative forces}
\title{Computing bulk and shear viscosities from simulations of fluids 
with dissipative and stochastic interactions}

\author{Gerhard Jung}
\email{jungge@uni-mainz.de}
\affiliation{Institut f\"ur Physik, Johannes Gutenberg-Universit\"at Mainz, 
Staudingerweg 9, D-55099 Mainz, Germany}
\author{Friederike Schmid}
\email{friederike.schmid@uni-mainz.de}
\affiliation{Institut f\"ur Physik, Johannes Gutenberg-Universit\"at Mainz, 
Staudingerweg 9, D-55099 Mainz, Germany}

\begin{abstract}

\noindent Exact values for bulk and shear viscosity are important
to characterize a fluid and they are a necessary input for a continuum description.
Here we present two novel methods to compute bulk viscosities by non-equilibrium molecular
dynamics (NEMD) simulations of steady-state systems with periodic boundary
conditions -- one based on frequent particle displacements and one based on the
application of external bulk forces with an inhomogeneous force profile. 

 In equilibrium simulations, viscosities can be determined from the stress tensor
fluctuations {\em via} Green-Kubo relations; however, the correct incorporation
of random and dissipative forces is not obvious. We discuss different
expressions proposed in the literature and test them at the example of a
dissipative particle dynamics (DPD) fluid.

%There exist various well-known methods to study the rheology of fluids
%that are interacting via conservative forces only. However, the correct 
%treatment of random and dissipative forces is still under debate.\\
%In this paper we present fully consistent results for both viscosities
%using Green-Kubo relations and non-equilibrium molecular dynamics
%(NEMD). Green-Kubo relations are calculated using the formulae derived
%by Ernst and Brito \cite{ernst}, while NEMD results were obtained using two novel
%techniques based on ideas of M\"uller-Plathe \cite{plathe} and Backer \cite{poiseuille}.

\end{abstract}

\maketitle

% ------------------------------------------------------------------------------------ %

\section{Introduction}

To describe the flow of an isotropic and compressible Newtonian fluid on a
continuum level, it is necessary to have precise values of both shear viscosity
$ \eta $ and bulk viscosity $ \zeta $. In a Newtonian fluid, the shear
viscosity $ \eta $ relates a shear flow $ \partial_z u_x$ to the off-diagonal
component $ \sigma_{xz} $ of the stress tensor and is thus a measure for the
resistance of a fluid element against \emph{deformation of shape}.
Similarly, the bulk viscosity $ \zeta $  relates a divergence in the flow field
$ \partial_\alpha u_\alpha $ to the trace of the stress tensor $ \Tr \sigma $
and is thus a measure for the resistance of a fluid element against
\emph{deformation of volume}. For many flows only shear viscosity is of
interest. However, bulk viscosity can play an important role for the
Brownian motion of a large particle (e.g. a colloid) or especially for
shock wave problems (e.g.  damping of a sound wave). In fact shockwave problems
are exactly the systems chosen experimentally to determine the bulk viscosity
\cite{schmidt}.

There exist two different ways to determine transport coefficients like shear
and bulk viscosity of a fluid using molecular dynamics (MD). One can create a
non-equilibrium state and calculate the transport coefficients using direct
measurements (force, stress or flow measurements) \cite{ashurst}. We will refer
to this as non-equilibrium molecular dynamics (NEMD). Alternatively,
equilibrium fluctuations are evaluated to determine transport coefficients
using Green-Kubo \cite{greenkubo} or Einstein-Helfand relations
\cite{helfand}.

In NEMD, it is favorable to simulate a non-equilibrium steady-state with
time-independent statistical properties. This is easily possible in the case of
steady-state shear flow: One can use boundary-driven shear flow \cite{schmid1},
Lees-Edwards boundary conditions \cite{leesedwards}, force-driven Poiseuille
flow \cite{poiseuille} or momentum interchange \cite{plathe}. All these methods
have been used to determine the shear viscosity $ \eta $ of fluids. However,
creating a steady-state divergent flow field to measure the bulk viscosity $
\zeta $ is much more complicated.  As mentioned above, the bulk viscosity is
related to the deformation of the volume of a fluid element and thus to a
change of the local thermodynamic state of the system. This is the reason why
it is difficult to measure the bulk viscosity in a steady-state experiment
\cite{hoover1}. To the best knowledge of the authors, the only NEMD
calculations of the bulk viscosity use either cyclic compression
\cite{hoover1,hoover2} or the relaxation of an instantaneous distortion
\cite{heyes, ciccotti}. The methods used in these studies cannot be applied
to systems with stochastic dynamics -- either because they are explicitly
designed for Hamiltonian systems \cite{hoover1,hoover2,ciccotti}, or, in the
case of Ref.\ \cite{heyes}, because they cannot be used to determine the
instantaneous contribution of the random force to the viscosity (as already
mentioned in \cite{heyes}) due to the finite time step in MD simulations.
{In addition to the above mentioned calculations, there have been extensive NEMD studies of the elongationaly viscosity using the so called SLLOD equation of motion \cite{elo1,elo2,elo3}. This boundary-driven NEMD method does not rely on a Hamiltonian dynamics and it should be possible to generalize it to stochastic equations of motion. However, in order to calculate the bulk viscosity, it will be necessary to choose a vanishing strain rate in order to minimize the change of volume.}
Therefore, all these methods depend on a perturbation parameter $
\epsilon $ that has to be chosen very small to get reliable results, hence they
can only be used in the linear response regime.

In that regime, using Green-Kubo or Einstein-Helfand relations is an
attractive alternative to NEMD methods. The great advantage is that one can
evaluate transport coefficients without having to create a non-equilibrium
state. In conservative systems, both relations are well-understood and
often used. However, the correct way to account for random and dissipative
forces is less clear. In 1995 Espa\~{n}ol \cite{espanol} suggested a generic
Green-Kubo form  
\begin{equation}
\mu = \frac{V}{k_\text{B} T} \int_{0}^{ \infty} dt \left \langle  I^C(0) I^C(t) 
  + I^D(0) I^D(t) \right \rangle_0,
\label{eq:espanol}
\end{equation} 
which depends on the conservative and dissipative projected momentum currents 
$ I^C(t) $ and $ I^D(t) $, respectively. {``Generic'' means, that the equation is not restricted to a specific viscosity. In Sec.~\ref{green-kubo} we will identify the projected momentum currents $ I(t) $ to find practical Green-Kubo relations for the shear viscosity $ \eta $ and the bulk viscosity $ \zeta $.} The formula implies that the random force has no direct contribution and the
total viscosity is just a summation of the conservative and the dissipative
contribution. More recently, Espa\~{n}ol and V\'{a}zquez \cite{espanol2002} 
and later Ernst and Brito \cite{ernst} suggested another generic Green-Kubo form,
\begin{equation}
 \mu = \mu_\infty + \frac{V}{k_\text{B} T} \int_{0}^{ \infty} dt \left \langle  
        (I_- e^{t \mathcal{L}} I_+ \right \rangle_0,
 \label{eq:ernst}
\end{equation} 
where the instantaneous viscosity $ \mu_\infty $ denotes a contribution of
stochastic forces. The time evolution is defined by the pseuodstreaming
operator $ e^{t \mathcal{L}} $ and $ I_\pm = I^C \pm I^D  $ (details in
Sec.~\ref{green-kubo} and Ref.~\cite{ernst}). This expression explicitly
accounts for the fact that the dissipative force is \emph{not invariant under
time reversal symmetry}. 
To our best knowledge, none of these formulae have
been verified by simulations in the presence of random and dissipative forces
to this date. 

Recently Espa\~{n}ol \cite{espanol2009} also presented a
generalized Einstein-Helfand form that can be used if the underlying dynamics
is dissipative and stochastic. It could be shown that both relations
should give the same result. In the present work we focus on generalized
Green-Kubo expressions that have the useful advantage of separating the
stochastic contribution and should therefore give better statistics. In
the future, it will also be interesting to test Einstein-Helfand relations in
the presence of stochastic and dissipative interactions.

The purpose of the present work is two-fold. First, we propose two novel
NEMD techniques to measure the bulk viscosity $\zeta$ from fluid simulations of
steady-state systems with periodic boundary conditions.  One, denoted
''particle transfer method'', creates a divergent flow field by manually
displacing particles. This method is most efficient in systems of particles
interacting by soft potentials. The other, denoted ''force driven method'',
makes use of a spatially varying body force and a non-zero center of mass
velocity and can be used in a wider range of molecular dynamics simulations.

Second, we test the Green-Kubo relations (\ref{eq:espanol}) and
(\ref{eq:ernst}) using the example of a dissipative particle dynamics (DPD) fluid
and compare the resulting values with the results from NEMD simulations.

Our paper is organized as follows. In Sec.~\ref{NEMD}, we present our
new NEMD methods for determining bulk viscosities from fluid simulations.
In Sec.~\ref{green-kubo}, we briefly introduce the different Green-Kubo 
relations for the shear and bulk viscosity. In Sec.~\ref{results}, we
present simulation results for DPD fluids and compare the values of the
viscosity parameters obtained with different methods. We summarize and
conclude in Sec.~\ref{summary}.

% ------------------------------------------------------------------------------------ %

\section{NEMD}
\label{NEMD}

In this section we present the NEMD methods for calculating viscosity
parameters from fluid simulations.  First we review the momentum interchange
method to create a steady-state shear flow that was introduced by
M\"uller-Plathe \cite{plathe}. Then we introduce our novel techniques to
create a steady-state divergence of the flow field. The last section explains
how to evaluate the local stress tensor to determine precise values of the
viscosities. The methods are illustrated by the example of DPD simulations
of fluids \emph{without conservative interactions}.

{In DPD the particles are interacting via dissipative
and random pair forces, which are constructed in a way that the total momentum
is conserved \cite{dpd}. Both forces are connected via fluctuation dissipation
theorems such that a proper canonical distribution is reached at
equilibrium \cite{espanoldpd}.  Therefore DPD is a Galilean invariant
thermostat and can be used to study hydrodynamics. In fact, Marsh \emph{et
al.} \cite{marsh} showed that DPD fulfills the hydrodynamic equations
(Navier-Stokes equation) and calculated theoretical values for transport
coefficients.}

{The DPD equations of motion can be written as stochastic 
differential equations \cite{espanoldpd}}

\begin{eqnarray}
\text{d} \mathbf{r}_i &= &\frac{\mathbf{p}_i}{m} \text{d}t \label{eq:dpdforce}\\
	\text{d} \mathbf{p}_i &=&  \sum\limits_{j\neq i}^{} -\gamma \omega_\text{D}(\mathbf{r}_{ij})(\mathbf{e}_{ij}\cdot\mathbf{v}_{ij})\mathbf{e}_{ij} \text{d}t +  \sum\limits_{j\neq i}^{} \sigma \omega_\text{R}(\mathbf{r}_{ij})\mathbf{e}_{ij} \text{d}W_{ij} \nonumber
\end{eqnarray}
{with the velocity difference $ \mathbf{v}_{ij} = \mathbf{v}_i - \mathbf{v}_j $, 
the distance $ \mathbf{r}_{ij} = \mathbf{r}_i - \mathbf{r}_j $ and the 
fluctuation dissipation theorems $ \sigma = \sqrt{2 k_\text{B} T \gamma} $ 
and $ \omega_\text{D}(r)=\omega_\text{R}(r)^2 $.}

{In the present work, we use the weight function
$\omega_\text{R}(r) = 1- \frac{r}{r_\text{cut}} $.
The simulation units are defined by setting $k_B T = 1$ (energy), 
$r_\text{cut} = 1$ (length) and 
$\tau = r_\text{cut} \sqrt{m/k_B T} = 1$ (time). 
In these units, we choose
the DPD parameter $ \gamma = 5 $ and the time step
$ \Delta t = 0.01 $.}

\subsection{Steady-state shear flow}
\label{ssec:shearflow}
If a shear flow $ \partial_z u_x $ emerges in a fluid, it will be
damped due to shear viscosity $ \eta $. This damping can be described by
a momentum flux
\begin{equation}
 j_z(p_x) = -\eta \frac{\partial u_x}{\partial z}  
\end{equation}
which transports momentum $ p_x $ in $ - z $-direction. To maintain the shear
flow it is therefore necessary to counteract the damping by enforcing
a similar momentum flux in $ +z $-direction.
\begin{figure}[h]
\includegraphics[]{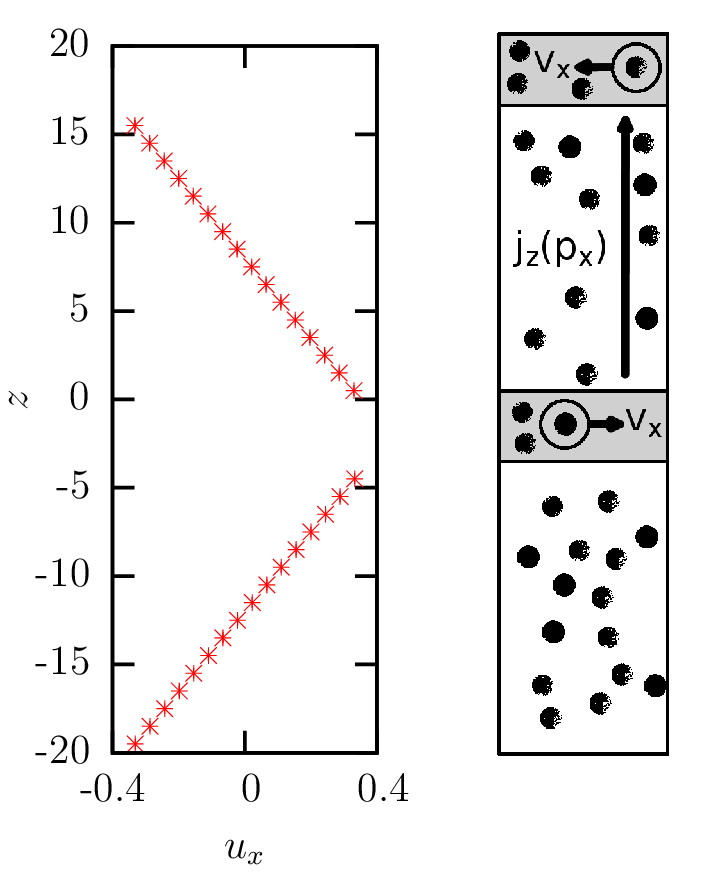}
\caption{\label{fig:shear_flow}Momentum interchange method to generate 
steady-state shear flow. Right panel illustrates the idea of the method.
Left panel shows the velocity profile from simulation results for a
DPD fluid at density $\bar{\rho} = 4$.}
\end{figure}
	
The idea of the momentum interchange method \cite{plathe} is to produce this
counter flux by swapping the momentum of particles with a certain rate. The
algorithm of one interchange works as follows (see Fig.~\ref{fig:shear_flow},
right):

\begin{enumerate}
\item Choose the particle with the highest velocity in $ +x $-direction 
in the middle slab
\item Choose the particle with the highest velocity in $ -x $-direction 
in the top slab
\item Swap the momentum of these two particles
\end{enumerate}

The shear rate $ \dot{\gamma} = \partial_z u_x $ will depend on the rate of
these momentum interchange steps and the velocity profile will be linear in
good approximation if the interchange rate is not too large (see
Fig.~\ref{fig:shear_flow}, left).

\subsection{Steady-state divergent flow}
\label{ssec:divflow}

Unfortunately, the momentum interchange method cannot be used for creating
steady-state {\em divergent} flow (with gradient $\partial_x u_x$). This is
because the transported momentum $ p_x $ and the direction of flux $
j_x(p_x) $ are no longer orthogonal. In fact, the directions of particle
and momentum transport are parallel and therefore a transport of {\em particles} is
necessary to maintain the steady-state.  This can be achieved in two different
ways: either by manually displacing particles or by imposing a
global flow. Our two methods of generating steady-state divergent flow are
based on these two types of mass transport. 

The methods can be motivated
theoretically by solving the mass and momentum continuity equations with
appropriate source terms. {The local conservation of mass on a continuum level with source term reads:}
\begin{equation}
\partial_t \rho + \nabla \cdot (\rho \mathbf{u}) = Q_\rho(\mathbf{r})
\end{equation}
{with density $ \rho(\mathbf{r},t) $, velocity field $ \mathbf{u}(\mathbf{r},t) $ and mass source term $ Q_\rho(\mathbf{r}) $.
Similarly one can write down the local conservation of momentum:}
\begin{equation}
\partial_t (\rho \mathbf{u}) 
   + \nabla (\rho \mathbf{u} \otimes \mathbf{u}) 
   + \nabla \sigma^\text{N} = Q_\rho(\mathbf{r}) \mathbf{u} 
   + \mathbf{f}(\mathbf{r})
\end{equation}
{with the external force field $ \mathbf{f}(\mathbf{r})$ and the constitutive relation \cite{hansen}:}
\begin{equation}
 {\sigma}^\text{N}(\mathbf{r},t) 
    = p\mathbf{I} + \left (\frac{2}{3}\eta
    -\zeta \right ) \nabla \cdot \mathbf{u}\mathbf{I} 
    - \eta \left( \nabla \mathbf{u}+ \nabla \mathbf{u}^T \right) 
 \label{eq:newtonstress}
\end{equation}
{with flow field $ \mathbf{u}(\mathbf{r},t) $, pressure $ p $, shear viscosity $
\eta $ and bulk viscosity $ \zeta $.}

{The external force and mass source terms can be chosen freely and allow us
to manipulate the flow and density profiles. However, to conserve global mass
and momentum both the mass source term $ Q_\rho(\mathbf{r}) $ and the force
field $ \mathbf{f}(\mathbf{r}) $ must fulfill the relations:}
\begin{eqnarray}
 &\int_V Q_\rho(\mathbf{r}) \text{d}\mathbf{r} = 0 \nonumber\\
 &\int_V \mathbf{u}(\mathbf{r})Q_\rho(\mathbf{r}) \text{d}\mathbf{r}
      +\int_V \mathbf{f}(\mathbf{r}) \text{d}\mathbf{r} = 0 \nonumber
\end{eqnarray}
{In the following, we assume that the fluid is barotropic, i.e., there
exists a unique relation $p=P(\rho)$.  We are interested in stationary
solutions $ \partial_t \rho=0 $ and $ \partial_t (\rho \mathbf{u}) = 0 $,
where $ \mathbf{u} = u \mathbf{e}_x $ exhibits a gradient in
$ x $-direction and profiles are constant in all other directions.
Assuming that higher order derivatives of the flow field can
be neglected, we obtain the following equations:}
\begin{eqnarray}
Q_\rho(x)&=&\partial_x (\rho u) \nonumber\\
f(x) &=& - Q_\rho(x) u +  P'(\rho) \partial_x \rho 
         + \partial_x ( \rho u^2)
\label{eq:profile}
\end{eqnarray}

\subsubsection{Particle transfer method}
\label{sssec:exchange}

\begin{figure}
	\includegraphics{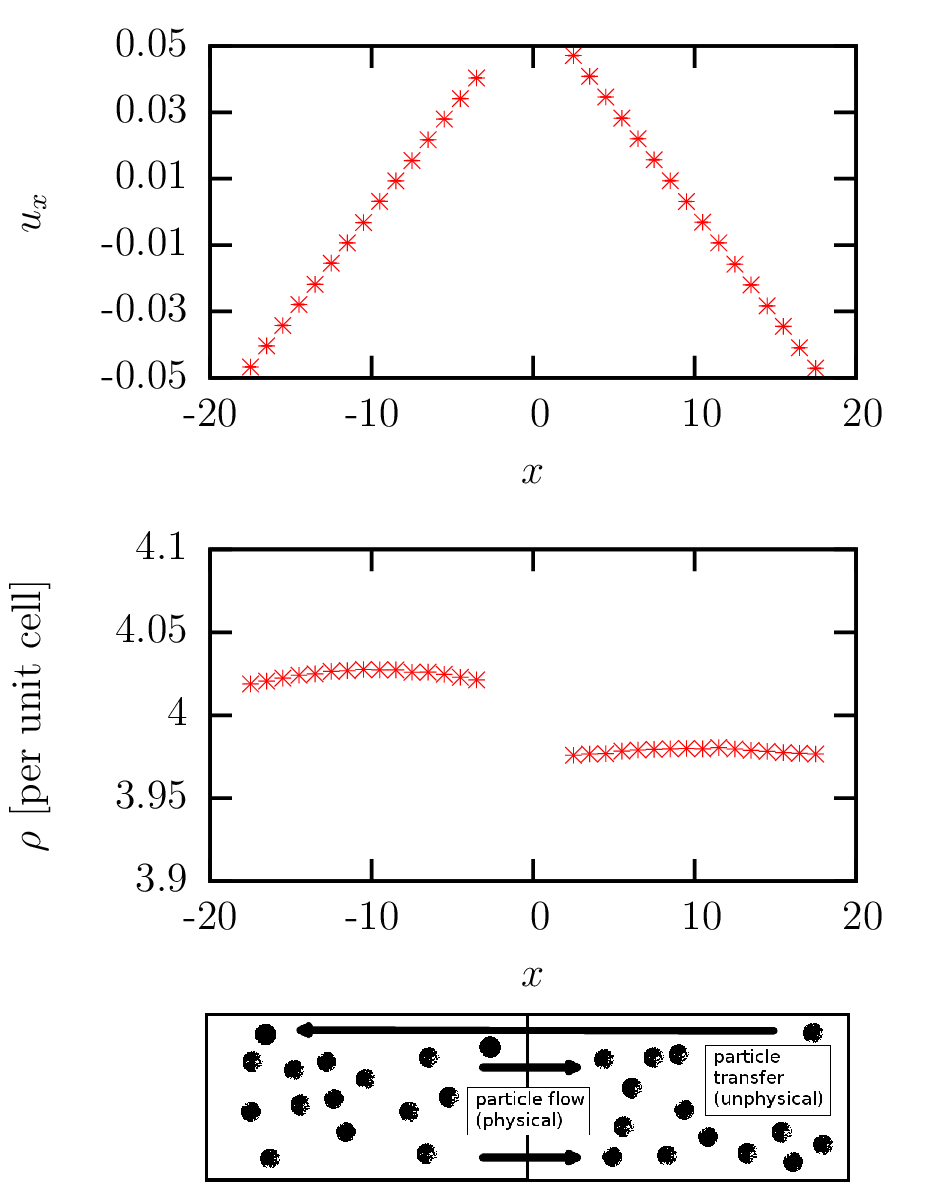}
	\caption{\label{fig:divergence_exchange}
Particle transfer method to generate steady-state divergent flow. Bottom
panel illustrates the idea of the method. Upper panels show corresponding
simulation results for the velocity (top) and the density (middle) of a DPD
fluid at mean density $\bar{\rho} = 4$. In this method, the density is roughly
constant.}
\end{figure}

{In the particle transfer method, we aim at creating
a velocity gradient while keeping the density profile 
constant, $ \rho(x) = \bar{\rho} $. This leads to}
\begin{eqnarray}
Q_\rho(x)&=&\bar{\rho} u'(x) \\
f(x) &=& \bar{\rho} \: u \: u'(x)
\end{eqnarray}
{To get a linear profile $u(x) = \epsilon \; \left| x \right|$, 
one therefore has to choose a mass source term}
\begin{equation}
Q_\rho(x)=\bar{\rho} \: \epsilon  \: \text{sign}(x)
\end{equation}
{and a force term $f(x) = \bar{\rho} x \epsilon^2 \approx 0$ which 
vanishes at order ${\cal O}(\epsilon)$.
These source terms can be realized by transferring
particles between two halves of the box} at a certain rate (see 
Fig.~\ref{fig:divergence_exchange}). The algorithm is very simple:

\begin{enumerate}
  \item Choose a random particle in the right half of the simulation box
  \item Place it at a random position in the left half of the simulation box
\end{enumerate}

This algorithm conserves momentum and -- in the special case of vanishing
conservative forces -- also energy. As shown in
Fig.~\ref{fig:divergence_exchange}, the resulting velocity profile is
approximately linear and the density profile almost constant, in perfect agreement with the above described theoretical prediction. Therefore, this algorithm is perfectly suitable to study
the bulk viscosity $ \zeta $ of fluids that are interacting only via soft
potentials.\\
\\
However, the particle transfer method can become problematic
in the presence of hard-core potentials, like the widely used 
Lennard-Jones potential, because particle insertion in dense
fluids is difficult and usually associated with large energy penalties.

\subsubsection{Force driven method}
\label{sssec:force}

{In the force driven method there is no particle transfer and hence no mass source term, $ Q_\rho(\mathbf{r})=0 $. This directly leads to $\rho u =$ const. A velocity gradient
is invariably associated with a density gradient, and can only exist
if the mean velocity $\bar{u}$ is nonzero.  }

\begin{figure}[]
	\includegraphics{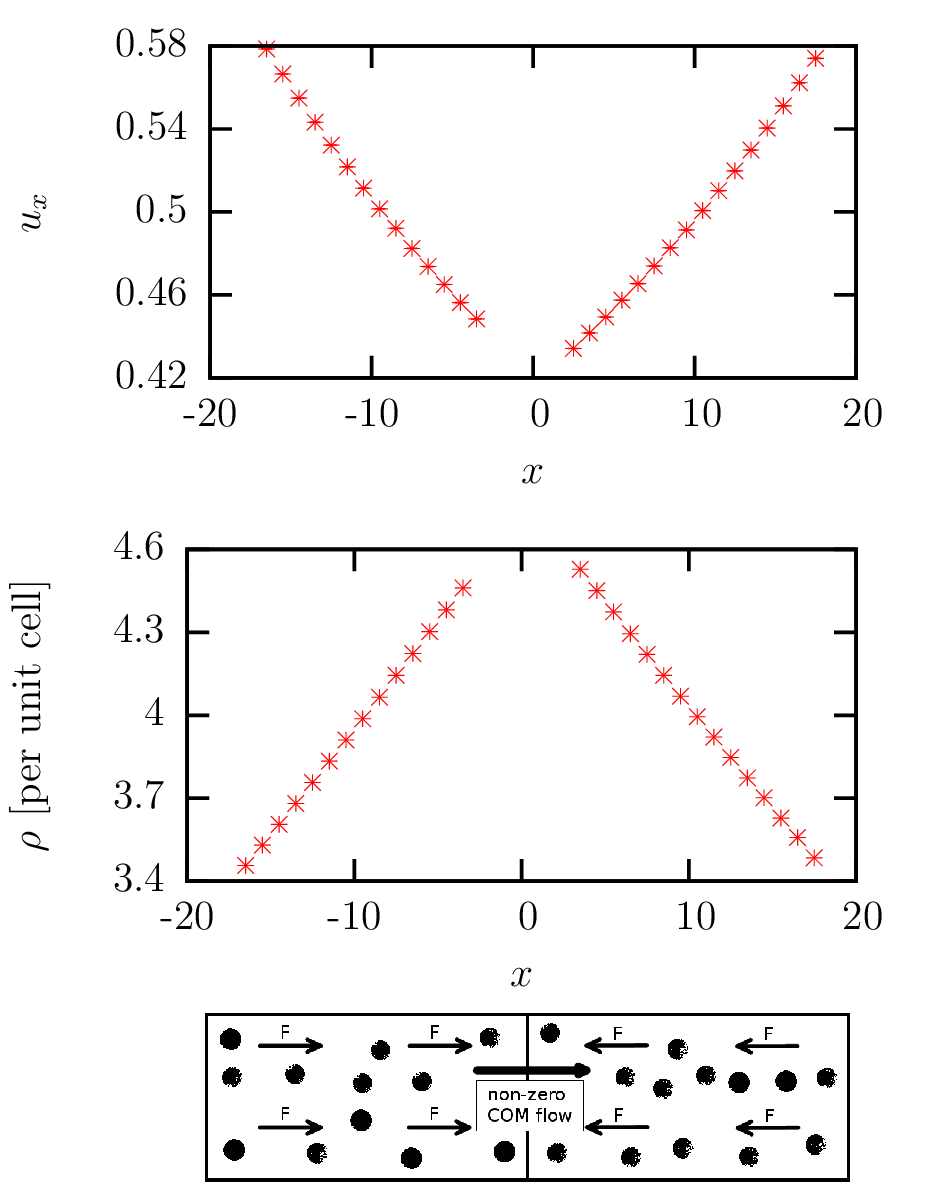}
	\caption{\label{fig:divergence_force}
Force-driven method to generate steady-state divergent flow.  Bottom panel
illustrates the idea of the method. Upper panels show corresponding simulation
results for the velocity (top) and the density (middle) of a DPD fluid at mean
density $\bar{\rho} = 4$. In this method, the density varies spatially.}
\end{figure}

\noindent{ The force density now reads}
\begin{equation}
f(x) = -\frac{\rho}{u}(P'(\rho)-u^2)\: u'(x)
\end{equation}
{To create a linear profile 
$ u(x) = \bar{u} + \epsilon \: \left| x \right|$,
one therefore has to apply an external force}
\begin{equation}
f(x) = - \epsilon \: C \: \text{sign}(x)
+ {\cal O}(\epsilon^2),
\end{equation}
{where the constant is given by $C = \left( P'(\bar{\rho}) \: \bar{\rho}/\bar{u}
- \bar{\rho} \: \bar{u} \right)$. \\
These equations explain how to create a linear velocity profile in the absence of a mass source term:} One has to create a steady-state particle flow in the presence of periodic boundary conditions by imposing a non-zero center of mass
velocity. To create a divergent flow field, one has to combine this
global background flow with an external force acting on all particles,
which changes sign between the two halves of the box (see
Fig.~\ref{fig:divergence_force}).

This method has the great advantage that one does not have to manually
change the position of particles. It can thus be used in combination
with hard-core potentials. It is also physically more ''realistic'' since 
it only requires an external force and a non-zero flow velocity. Therefore, 
the basic idea may be applicable in experiments.

The main disadvantage of this technique is that the gradient in the flow
field is unavoidably associated with a density gradient. This is a problem because the bulk viscosity strongly
depends on the density. One can reduce the problem by applying a very
small force. However, very long simulations are then necessary to obtain
sufficiently good statistics. The solution used in this work is to calculate a
bulk viscosity $ \zeta $ for every bin in the simulation box and associate it
to the density in the respective box. In this way one obtains many data
points for various densities that can be used to determine not only the bulk
viscosity for a constant density but also the density dependence (see
Sec.~\ref{ssec:resultsbulk}).

\subsection{Local stress tensor}
\label{localstress}

We calculate the viscosity by direct evaluation of the local stress
tensor using both Newton's constitutive relation and a localized
Irving-Kirkwood formula. This has the great advantage that one can decouple the
problems of creating a steady-state flow and calculating viscosities. When
using this method it is also possible to distinguish between the dissipative
and the conservative contribution to the viscosity.

\subsubsection{Newton's constitutive relation}
\label{constitutive}

The local stress tensor of an isotropic, compressible Newtonian fluid can be
calculated using Newton's constitutive relation (see Eq.~(\ref{eq:newtonstress})).

It can be evaluated in a particle simulation by dividing the simulation box
into bins and calculating the flow field (mean velocity of all particles) in
each bin. The errors due to discretization are small because of the linearity
of the observed flow profiles.

\subsubsection{Irving-Kirkwood formula}
\label{irvingkirkwood}

In a particle simulation the global stress tensor can also be calculated using
the Irving-Kirkwood formula \cite{irving}:
\begin{eqnarray}
\label{eq:irving}
&\sigma^\text{IK}_{\alpha \beta} 
    = \left\langle \sigma^C_{\alpha \beta}(t) 
      + \sigma^D_{\alpha \beta}(t) 
      + \sigma^R_{\alpha \beta}(t) \right\rangle  \\
 &\sigma^C_{\alpha \beta}(t)
      = \frac{1}{V} \sum\limits_{i<j}^{} r_{ij\alpha}(t) F^C_{ij\beta}(t) 
        + \frac{1}{V}\sum\limits_{i}^{} m u_{i \alpha}(t) u_{i \beta}(t) \nonumber \\
 &\sigma^D_{\alpha \beta}(t)
      = \frac{1}{V}\sum\limits_{i<j}^{} r_{ij\alpha}(t) F^D_{ij\beta}(t) \nonumber\\
 &\sigma^R_{\alpha \beta}(t)
      =\frac{1}{V}\sum\limits_{i<j}^{} r_{ij\alpha}(t) F^R_{ij\beta}(t)  \nonumber
\end{eqnarray}
with distance $ \mathbf{r}_{ij} = \mathbf{r}_i - \mathbf{r}_j $,
conservative force $ \mathbf{F}^{\text{C}}_{ij} $, dissipative force $
\mathbf{F}^{\text{D}}_{ij} $ and random force $ \mathbf{F}^{\text{R}}_{ij} $
between two particles. In our model system, there are no conservative
forces and {the kinetic contribution is equal to the conservative stress tensor}. This equation can be
evaluated locally in a homogeneous fluid by considering the contribution of all
particles in one bin only.  

\subsubsection{Calculating viscosities}
\label{ssec:calcviscosity}
Evaluating the local stress tensor using the aforementioned formulae in a
steady-state shear flow makes it possible to calculate the shear viscosity:
\begin{equation}
 \eta =  \frac{\sigma^\text{IK}_{xz}(\mathbf{r})}{\partial_z u_x (\mathbf{r})} 
\end{equation}

Similarly, in a steady-state divergent flow field, the bulk viscosity
can be calculated {\em via}
\begin{equation}
\zeta =  \frac{1}{3}\frac{\sigma^\text{IK}_{xx}(\mathbf{r}) 
  + \sigma^\text{IK}_{yy}(\mathbf{r})
  + \sigma^\text{IK}_{zz}(\mathbf{r}) - 3 p}{\partial_x u_x (\mathbf{r})}
\label{eq:viscosity}
\end{equation}

\begin{figure}[t]
\includegraphics{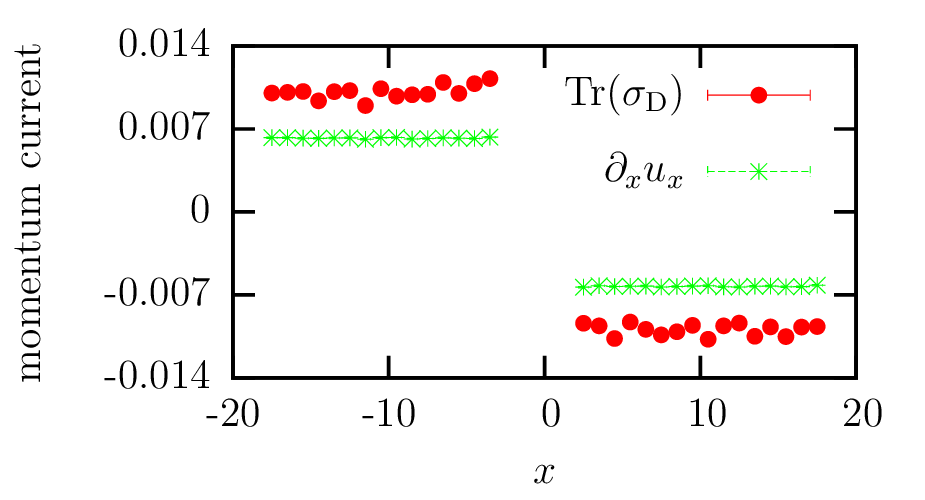}
\caption{\label{fig:divandstress} 
Profiles of divergence of the flow field and trace of the 
dissipative contribution to the stress tensor, obtained
with the particle transfer method for a DPD fluid with
density $ \bar{\rho} = 4.0 $.}
\end{figure}

The procedure of calculating the bulk viscosity is demonstrated in 
Fig.~\ref{fig:divandstress} using the example of the particle transfer method
(see Sec.\ref{sssec:exchange}). Both the flow field and the stress tensor
are evaluated locally and used to determine the bulk viscosity $ \zeta $ in
each bin. The final viscosity is calculated using the
average of the viscosities.

% ------------------------------------------------------------------------------------ %

\section{Green-Kubo}
\label{green-kubo}

Green-Kubo formulae use equilibrium fluctuations of the projected momentum
current $ I_{\alpha \beta} $ to calculate transport coefficients (see
Eqs.~(\ref{eq:espanol}) and (\ref{eq:ernst})). The projected momentum
current is the projection of the instantaneous momentum current
fluctuation $ (\sigma_{\alpha \beta}(t) - \left \langle \sigma_{\alpha
\beta}\right \rangle_0) $ (see Eq.~(\ref{eq:irving})) on the irrelevant
hydrodynamic variables, i.e., on the subspace of variables that is
orthogonal to the slow variables of the hydrodynamic equations: energy, mass
and momentum density.  Here and in the following, the notation $\langle
\cdot \rangle_0$ refers to equilibrium averages.
For details see Ref.~\cite{ernst1974}).

\subsection{Shear viscosity}

{The shear viscosity $ \eta $ is related to the off-diagonal $ I_{xz} $ of the projected momentum current.
In this case, finding the projection is simple.} The mean values 
$\left \langle \sigma_{x z} \right \rangle_0$ are zero and the momentum
current is orthogonal to the relevant subspace:  
\begin{equation}
I_{xz} = \mathcal{P}_\perp (
\sigma_{xz} - \left \langle \sigma_{xz} \right \rangle_0)  = \sigma_{xz}   
\end{equation}
{With this relation we can find the Green-Kubo formulae for the shear viscosity $ \eta $.}
{Equation~(\ref{eq:shear_esp}) corresponds to the generic formula (\ref{eq:espanol}):}
\begin{eqnarray}
\label{eq:shear_esp}
 \eta = \frac{ V}{k_\text{B}T} \int_{0}^{ \infty} dt \left \langle
\sigma^C_{xz}(0)\sigma^C_{xz}(t)+\sigma^D_{xz}(0)\sigma^D_{xz}(t)
\right \rangle_0 \qquad
\end{eqnarray}
{and Equation~(\ref{eq:shear_ernst}) corresponds to the generic formula (\ref{eq:ernst}):}
\begin{eqnarray}
\label{eq:shear_ernst}
\nonumber
 \eta = &\eta_\infty& + \frac{ V}{k_\text{B}T} \int_{0}^{ \infty} dt \left \langle
(\sigma^C_{xz}(0)-\sigma^D_{xz}(0) (\sigma^C_{xz}(t)+\sigma^D_{xz}(t) )
\right \rangle_0\\
&\eta_\infty& 
  = \frac{ V}{k_\text{B}T}\frac{\Delta t}{2}  \left \langle {\sigma^R_{xz}}^2 \right \rangle_0,
\end{eqnarray}
{where the different contributions to the stress tensor $ \sigma_{xz} $ are defined as in Eq.~(\ref{eq:irving}).}

\subsection{Bulk viscosity} \label{greenbulk}

{The bulk viscosity $ \zeta $ is related to the diagonal $ I_{\alpha \alpha} $ of the projected momentum current.
Therefore, the situation is slightly more
complicated.} First, the mean currents $\left \langle \sigma_{\alpha \alpha}
\right \rangle_0 $ do not vanish.  Second, the momentum current
fluctuations are not orthogonal to the relevant subspace. This becomes
clear when considering the kinetic contribution $ \sigma^C_{\alpha
\alpha}(t) $, which is proportional to the kinetic energy and therefore
depends on the energy density -- a relevant variable. Therefore, the 
energy fluctuations enter the projected momentum current explicitly:

\begin{equation}
\label{eq:identify}
I_{\alpha \alpha} =
   \mathcal{P}_\perp ( \sigma_{\alpha \alpha} 
   - \left \langle \sigma_{\alpha \alpha} \right \rangle_0)  
= \sigma_{\alpha \alpha} 
   - \left \langle \sigma_{\alpha \alpha} \right \rangle_0 
   - \frac{1}{V} \frac{\partial p}{\partial e}
    (\mathcal{H} - \left\langle \mathcal{H}\right\rangle_0 ) 
\end{equation} 
with the average energy density 
$ e =  V^{-1} \left\langle \mathcal{H}\right\rangle_0 $.\\
 {This leads to the projected momentum currents:}
\begin{eqnarray}
\label{eq:momcur}
\nonumber
I_\zeta^C(t) &=&  \sum_{\alpha} \left[ \sigma^C_{\alpha \alpha}(t) 
  - \left \langle \sigma^C_{\alpha \alpha} \right \rangle_0 
  - \frac{1}{V} \frac{\partial p}{\partial e} (\mathcal{H}(t) 
  - \left\langle \mathcal{H}\right\rangle_0 ) \right]
\\
I_\zeta^D(t) &=&  \sum_{\alpha} \left[ \sigma^D_{\alpha \alpha}(t) 
  - \left \langle \sigma^D_{\alpha \alpha} \right \rangle_0 \right]
\end{eqnarray}
{With Eqs.~(\ref{eq:identify}), (\ref{eq:momcur}) and the considerations above, we are able to find the Green-Kubo formulae for the bulk viscosity $ \zeta $.}
{Equation~(\ref{eq:bulk_esp}) corresponds to the generic formula (\ref{eq:espanol}):}
\begin{eqnarray}
\label{eq:bulk_esp}
 \zeta =  \frac{ V}{k_\text{B}T} \int_{0}^{ \infty} dt \left \langle
I_\zeta^C(0)I_\zeta^C(t)+I_\zeta^D(0)I_\zeta^D(t)
\right \rangle_0 \qquad
\end{eqnarray}
{And Equation~(\ref{eq:bulk_ernst}) corresponds to the generic formula (\ref{eq:ernst}):}
\begin{eqnarray}
\label{eq:bulk_ernst}
\nonumber
 \zeta = &\zeta_\infty& + \frac{ V}{k_\text{B}T} \int_{0}^{ \infty} dt \left \langle
(I_\zeta^C(0)-I_\zeta^D(0))(I_\zeta^C(t)+I_\zeta^D(t) )
\right \rangle_0\\
&\zeta_\infty& 
  = \frac{ V}{k_\text{B}T} \frac{\Delta t}{18} 
    \left \langle [\sum_\alpha\sigma^R_{\alpha \alpha}]^2 \right \rangle_0 
\end{eqnarray}

For an ideal gas (like our DPD fluid without conservative forces) 
it is easy to calculate the thermodynamic relation: 
\begin{equation} \frac{\partial p}{\partial e} = \frac{2}{3} 
\end{equation}
This leads to a vanishing conservative contribution to the projected momentum current 
$ I_\zeta^C(t) = 0 $ .

% ----------------------------------------------------------------------------------------------------------------------- %

% ------------------------------------------------------------------------------------------------------------------------ %

\section{Results}
\label{results}

\subsection{Shear viscosity}
\label{ssec:resultshear}

In this section, we focus on the test of the Green-Kubo formulae
introduced in Sec.~\ref{green-kubo}. 

For comparison we first calculated the shear viscosity using different NEMD
methods. The first two are based on the momentum interchange method
(see Sec.~\ref{ssec:shearflow}), where we computed the viscosity both
by measuring the momentum flux (as suggested in Ref.~\cite{plathe}) and
by comparing local stress tensors (as discussed in Sec.~\ref{ssec:calcviscosity}). As a third, independent NEMD approach,
we also performed simulations of confined slabs with tunable-slip
boundaries at the walls (see Ref.~\cite{schmid1}) and determined
the viscosity by evaluating the force on the boundary that is needed 
to maintain the shear flow.

\begin{figure}
	\includegraphics{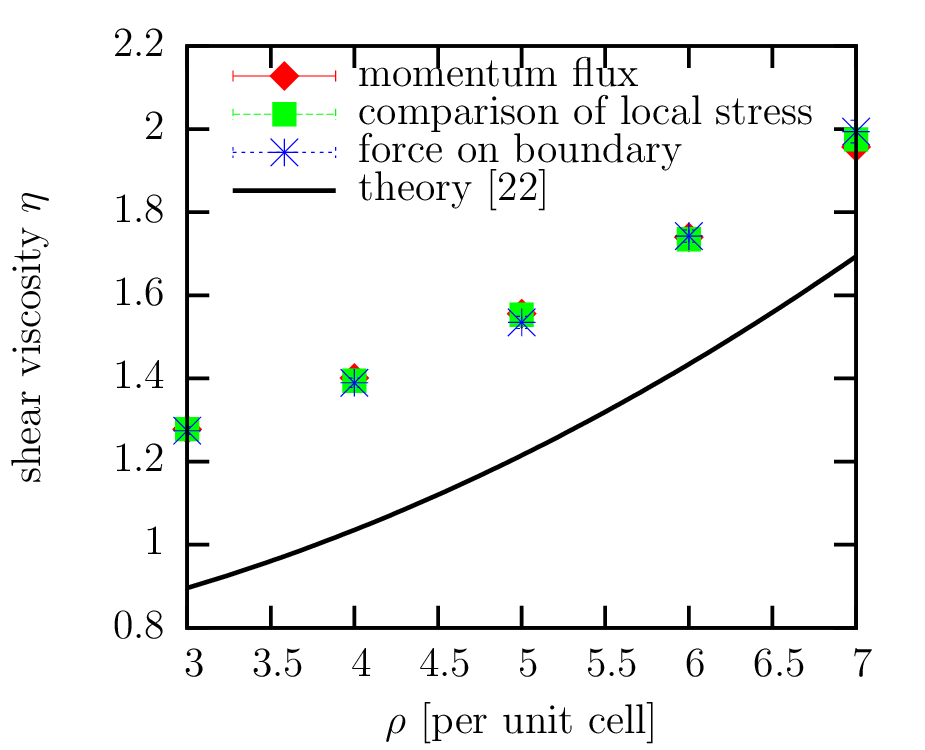}
	\caption{\label{fig:shear_nemd} 
  Results for the shear viscosity $ \eta $ for different densities 
  using several NEMD techniques (see Sec.~\ref{ssec:resultshear}) 
  and theoretical prediction. }
\end{figure}

As shown in Fig.~\ref{fig:shear_nemd}, all these different NEMD methods
give the same result for the shear viscosity. We therefore have a
reliable dataset to test the Green-Kubo formulae.  When comparing the
results to the theoretical prediction (see \cite{marsh}) one can observe a
constant offset of about 0.30. Hence, the $ \rho^2 $ dependence  of the
dissipative contribution $ \eta_\text{D} $ matches quite well but the kinetic
contribution $ \eta_\text{C} $ seems to be underestimated by the theory. This
interpretation is confirmed if we compare separately the dissipative 
and conservative contributions to the theoretical expression for the
shear viscosity with the corresponding NEMD results using the local stress
tensor. This discrepancy between simulations and theory has already
been noticed by Marsh \emph{et. al} in Ref.~\cite{marsh}. 

The Green-Kubo relations were evaluated by calculating the stress
autocorrelation functions with resolution $\Delta t$. Then the correlation
functions were integrated numerically using the trapezoidal rule, until the correlations
reached about 1~\% of their initial values. The tail was incorporated by fitting
the correlation function with the {power law} $ c(t) = A t^{-B} $ and
integrating it analytically.

\begin{figure}
	\includegraphics{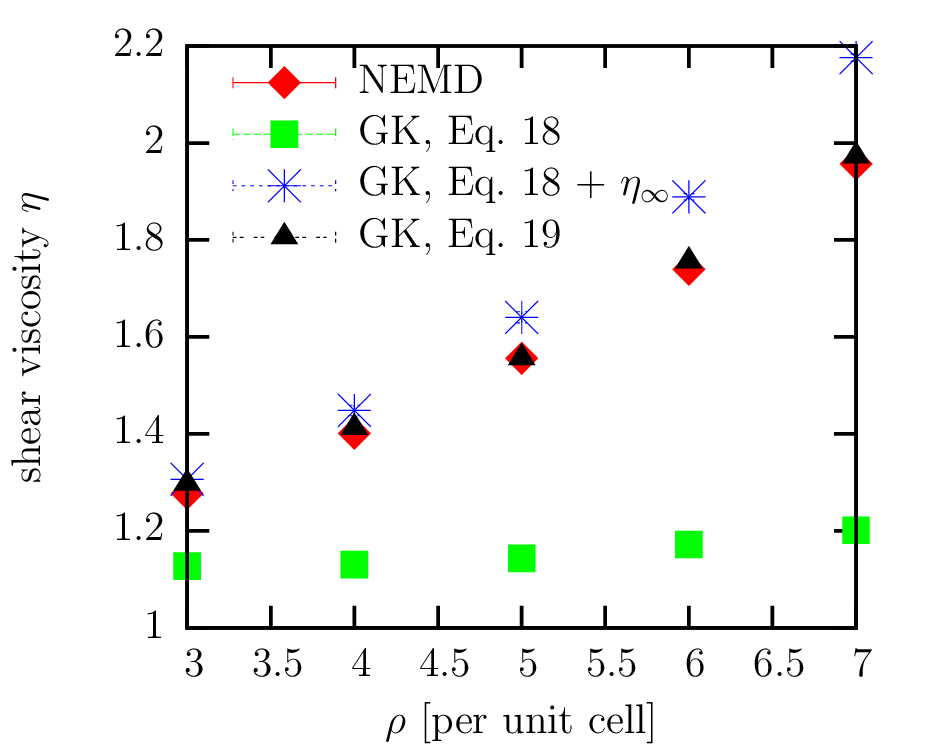}
	\caption{\label{fig:shear_gk} 
  Results for the shear viscosity $ \eta $ for different densities 
   using different Green-Kubo relations. The values are compared 
   to the NEMD results from the momentum flux technique.}
\end{figure}

The results of these calculations can be found in Fig.~\ref{fig:shear_gk}.
We used the momentum flux values for comparison with NEMD because 
they have the smallest statistical error.

The results obtained with the Green-Kubo formula (\ref{eq:shear_ernst}) are in
excellent agreement with the NEMD results. Good agreement is also obtained
with Eq.~(\ref{eq:shear_esp}) for small densities, if the instantaneous viscosity
$ \eta_\infty$ is added. The reason is that the dissipative contribution to
the viscosity is small at small densities. However, at large densities, the
results obtained with Eq.~(\ref{eq:shear_esp}) and the NEMD results significantly differ from
each other.

\subsection{Bulk viscosity}
\label{ssec:resultsbulk}
The bulk viscosity was determined using the NEMD method described in
Sec.~\ref{ssec:divflow} and the Green-Kubo formula~(\ref{eq:bulk_ernst}). The
results are also compared to theory \cite{marsh}.

\begin{figure}[]
\includegraphics{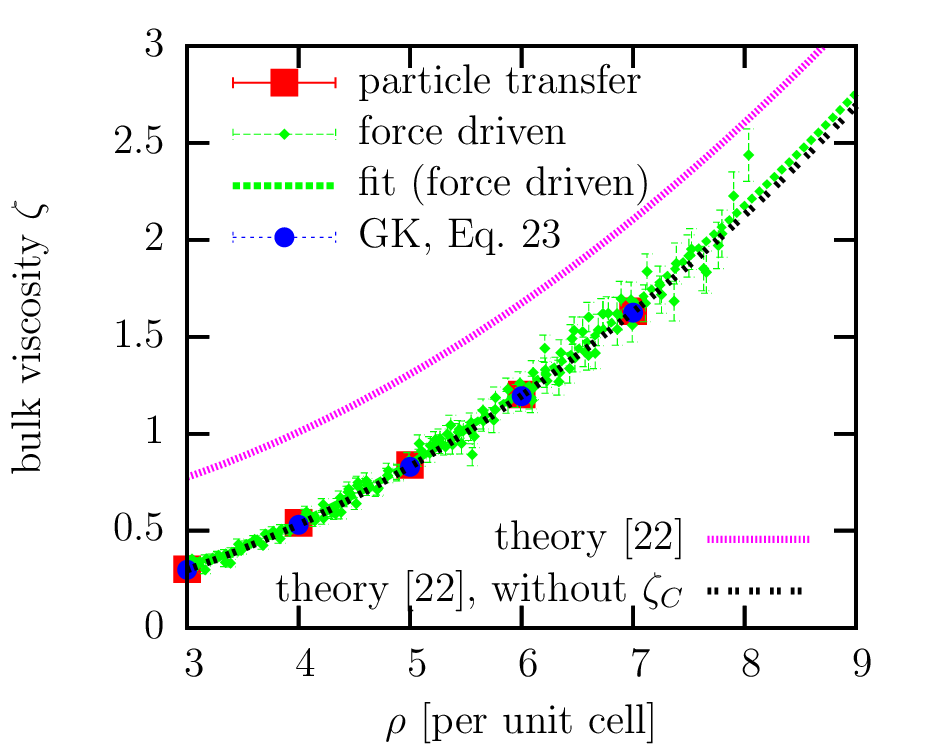}
	\caption{\label{fig:bulk} 
Results for the bulk viscosity $ \zeta $ for different densities 
using NEMD and Green-Kubo relations.}
\end{figure}

As in the case of the shear viscosity the agreement between the NEMD results
and the Green-Kubo relations is excellent.  However, the
simulations are not consistent with the theoretical prediction in
Ref.~\cite{marsh}. As shown in Fig.~\ref{fig:bulk} the discrepancy between
theory and simulation can be resolved by neglecting the kinetic contribution in
the theoretical prediction.

We rationalize the discrepancy between simulation and theory as follows:
The kinetic contribution in the theoretical prediction for the stress
tensor reflects the contribution of energy fluctuations to the pressure $p$,
and not to the viscous stress tensor. Hence it should not be associated with
the bulk viscosity $ \zeta $.

{In the projected momentum current Eq.~(\ref{eq:identify})}, the
corresponding term had to be subtracted. Likewise, it should also be subtracted
in the Chapman-Enskog expansion of the theoretical prediction
\cite{marsh}.

%\fs{It is also worth mentioning, that the relative statistical errors of the
%reported NEMD bulk viscosities are about 0.01 without a huge computational
%effort (a few days on a single core).}
%
% -----------------------------------------------------------------------------------------------------------------------------------
% %

\section{Summary and outlook}
\label{summary}

In the present paper, we have proposed two different techniques to create
a divergence in the flow field of a homogeneous fluid. Both methods use very
different mechanisms to create the flow. Therefore, one can choose the method
most suited for the system to investigate. It will be interesting to test these
techniques for other systems (e.g. an isothermal Lennard-Jones system) with
high densities and hard-core potentials.

Furthermore, we have proposed a way to calculate the viscosities by
comparing different expressions for the local stress tensor
(Sec.~\ref{localstress}), which is independent of the methods used to
create a non-equilibrium steady-state and can therefore be applied generally in
the presence of arbitrary steady-state flows. The disadvantage of this method
is the necessity to calculate the local stress tensor. While this task is
simple in the presence of short-range two-body potentials in a homogeneous
confinement, it can be more challenging for more complicated systems (e.g. in
the presence of charges).

The NEMD methods proposed here can be used to compute bulk
viscosities both in Hamiltonian systems and in dissipative systems
(with local momentum conservation). Moreover, they are not
restricted to the linear response regime, but can also be used 
to study nonlinear behavior.

Using NEMD simulations to determine the shear viscosity $ \eta $, we were
also able to test Green-Kubo relations in the presence of random and
dissipative forces. The results clearly support the validity of the generic
expression Eq.~(\ref{eq:ernst}), which includes the contribution of
stochastic forces and accounts for the lack of time reversal symmetry
in the dissipative force.

% ----------------------------------------------------------------------------------------------------------------------------------- %

\section*{Acknowledgment}

This work was funded by the German Science Foundation within 
SFB TRR 146. Computations were carried out on the Mogon Computing
Cluster at ZDV Mainz.

% --------------------------------------------------------------------------------------------------------------------------------- %

\appendix

% --------------------------------------------------------------------------------------------------------------------------------- %


\begin{thebibliography}{}
\bibitem{schmidt} A.J.~Schmidt et al., Appl.~Phys.~Lett. 92, 244107 (2008)
\bibitem{ashurst} W.T.~Ashurst and W.G.~Hoover, Phys.~Rev.~Lett. 31, 206 (1973)
\bibitem{greenkubo} M.S.~Green, J.~Chem.~Phys. 22, 398 (1954); R.~Kubo, J.~Phys.~Soc.~Jpn. 12, 570 (1957)
\bibitem{helfand} E.~Helfand, Phys.~Rev.~E 119, 1 (1960)
\bibitem{schmid1} J.~Zhou, J.~Smiatek, E.S.~Asmolov, O.I.~Vinogradova and F.~Schmid, Springer~HPC~'14, 19 (2014)
\bibitem{leesedwards} A.W.~Lees and S.F.~Edwards, J.~Phys.~C 5, 1921 (1972)
\bibitem{poiseuille} J.~Backer, C.~Lowe, H.~Hoefsloot and P.~Iedema, J.~Chem.~Phys. 122, 154503 (2005)
\bibitem{plathe} Florian~M\"uller-Plathe, Phys.~Rev.~E 59, 4894 (1998)
\bibitem{hoover1} W.G.~Hoover, A.J.C.~Ladd, R.B.~Hickman and B.L.~Holian, Phys.~Rev.~A 21, 1756 (1980)
\bibitem{hoover2} W.G.~Hoover, D.J.~Evans, R.B.~Hickman, A.J.C.~Ladd, W.T.~Ashurst and B.~Moran, Phys.~Rev.~A 22, 1690 (1980)
\bibitem{heyes} D.~Heyes, J.~Chem.~Soc., Faraday Trans. 2 80, 1363 (1984)
\bibitem{ciccotti} P.L.~Palla, C.~Pierleoni and G.~Ciccotti, Phys.~Rev.~E 78, 021204 (2008)
\bibitem{elo1} B.D.~Todd and P.J.~Daivis, J.~Chem~Phys. 107, 1617 (1997)
\bibitem{elo2} B.D.~Todd and P.J.~Daivis, Com.~Phys.~Comm. 177, 191 (1999)
\bibitem{elo3} A.~Baranyai and P.T.~Cummings, J.~Chem.~Phys. 110, 42 (1999)
\bibitem{espanol} P.~Espa\~{n}ol, Phys.~Rev.~E~52, 1734 (1995)
\bibitem{espanol2002} P.~Espa\~{n}ol and F.~V\'{a}zquez, Phil.~Trans.~R.~Soc.~A 360, 383 (2002)
\bibitem{ernst} M.H.~Ernst and R.~Brito, Europhys.~Lett. 73, 183 (2006)
\bibitem{espanol2009} P.~Espa\~{n}ol, Phys.~Rev.~E 80, 061113 (2009)
\bibitem{dpd} P.J.~Hoogerbrugge and J.M.V.A.~Koelman, Europhys.~Lett. 19, 155 (1992)
\bibitem{espanoldpd} P.~Espa\~{n}ol and P.~Warren, Europhys.~Lett. 30, 191 (1995)
\bibitem{marsh} C.A.Marsh, G.~Backx and M.H.~Ernst, Phys.~Rev.~E 56, 1676 (1997)
\bibitem{hansen} J.P.~Hansen and I.R.~McDonald, ``Theory of simple liquids'', Academic Press (2006)
\bibitem{irving} J.H.~Irving and J.G.~Kirkwood, J.~Chem.~Phys. 18, 817 (1950)
\bibitem{ernst1974} M.H.~Ernst and J.R.~Dorfman, J.~Stat.~Phys. 12, 311 (1974)
\end{thebibliography}
\end{document}